\documentclass[a4paper,11pt]{article}
\usepackage{pos}
\usepackage{lineno}

\title{Soft QCD results from ALICE}
\ShortTitle{Soft QCD results from ALICE}

\author*{Sushanta Tripathy (for the ALICE collaboration)}

\affiliation{INFN - sezione di Bologna,\\
   via Irnerio 46, 40126 Bologna BO, Italy}
   
\emailAdd{sushanta.tripathy@cern.ch}

\abstract{High-energy hadronic collisions are dominated by soft interactions with low momentum transfers. The description of these processes typically relies on phenomenological models. The soft QCD measurements, such as characterisation of the underlying event (UE) and study of light flavored hadrons, provide important constraints on the model parameters. They also provide additional insights into the recent measurements at the LHC where features normally attributed to QGP formation in Pb--Pb collisions have been observed even in pp and p--Pb collisions. Here, we present an analysis based on UE measurements applied to p--Pb collisions for the first time at the LHC to test the similarities between pp and p--Pb collisions. Furthermore, it is shown that the UE at midrapidity is correlated with the energy measured at forward rapidity by ALICE zero degree calorimeters (ZDC). Using ZDC, a multi-differential study is shown to disentangle initial and final-state effects on strange hadron production in pp collisions.}

\FullConference{%
 
 The Ninth Annual Conference on Large Hadron Collider Physics - LHCP2021\\
 7-12 June 2021\\
 Online
}

\begin{document}
\maketitle

\section{Introduction}
Most of the processes in high-energy hadronic collisions are dominated by soft interactions with low momentum transfers. The description of these processes using fundamental theory of strong interactions, quantum chromodynamics (QCD), is difficult due to their non-perturbative nature. Thus, the description of these processes typically relies on phenomenological models implemented in Monte-Carlo (MC) event generators. The soft QCD measurements, such as underlying event, light flavor production etc., provide important constraints on the model parameters and they also help in understanding the recent measurements at the LHC~\cite{ALICE:2017jyt,CMS:2010ifv,ALICE:2018pal} where features normally attributed to QGP formation in Pb--Pb collisions have been observed even in pp and p--Pb collisions. Here, we report recent results on soft QCD observables using the ALICE detectors at the LHC. The results on the underlying event at midrapidity and its correlation with the energy measured at forward rapidity are presented. Also, a multi-differential approach is presented using two forward detectors, {\it i.e.,} zero degree calorimeters (ZDC) and the V0 detectors placed on both sides of the interaction point of ALICE~\cite{ALICE:2008ngc,ALICE:2014sbx}, to explore the strangeness enhancement in pp collisions.

\section{Underlying-event measurements in pp and p--Pb collisions at $\sqrt{s_{\rm NN}}$ = 5.02\,TeV}
\begin{figure}[ht!]
\centering
\includegraphics[scale=0.35]{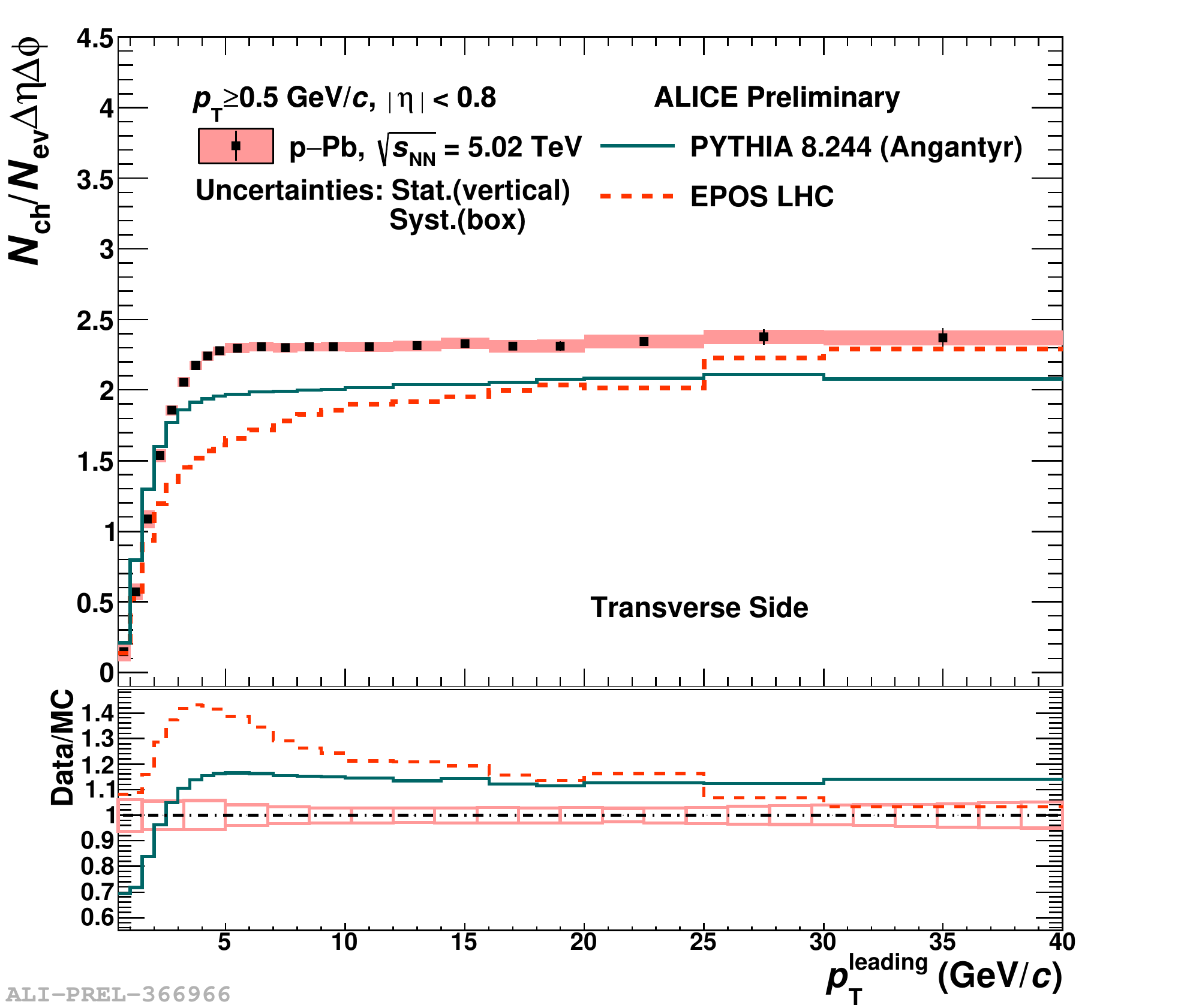}
\includegraphics[scale=0.35]{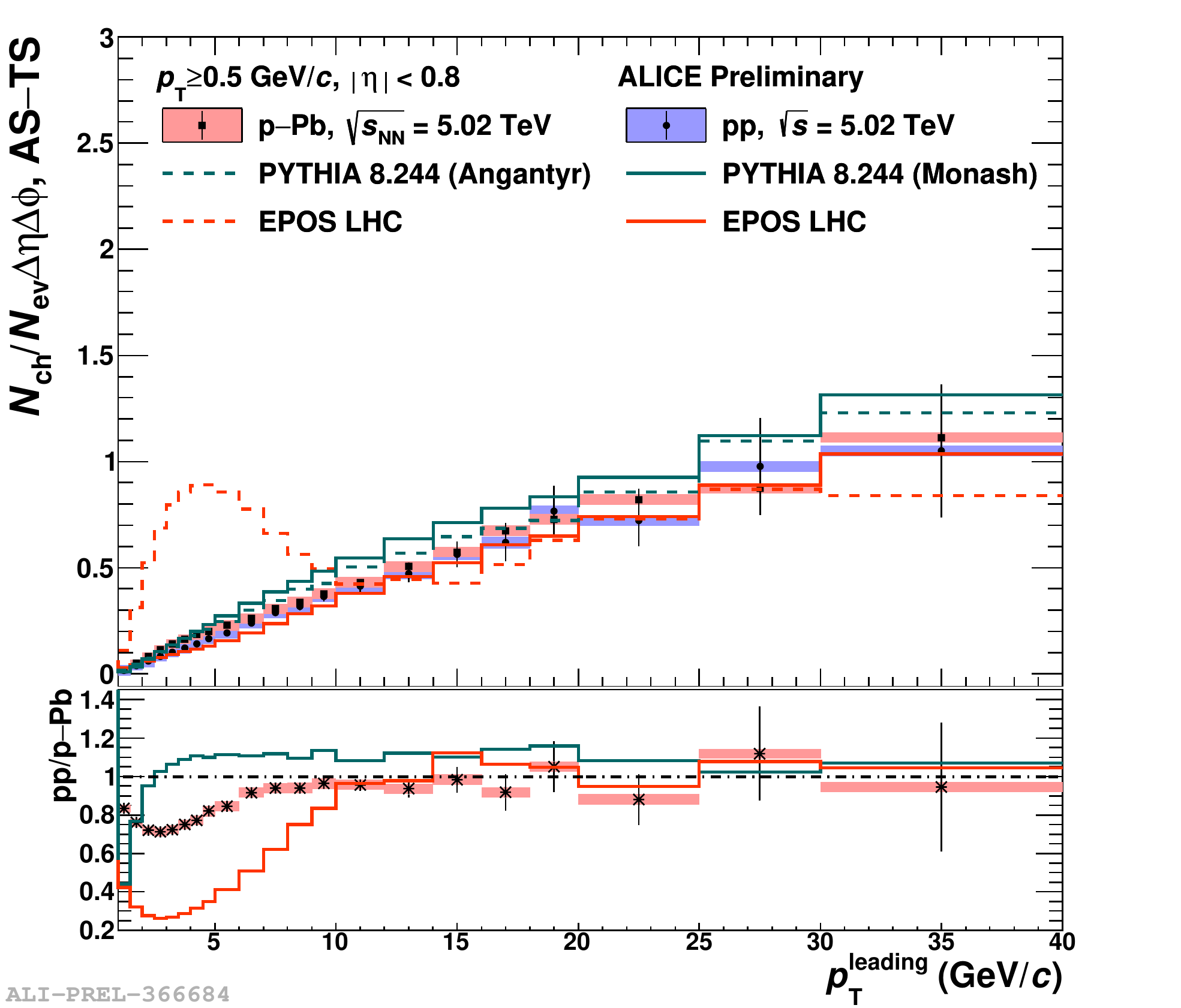}
\caption{Left: charged-particle density as a function of $p_{\rm T}^{\rm{leading}}$ for the transverse side in p--Pb collisions at $\sqrt{s_{\rm NN}}$ = 5.02\,TeV, Right: charged-particle density as a function of $p_{\rm T}^{\rm{leading}}$ from (away $-$ transverse) side for pp and p--Pb collisions at $\sqrt{s_{\rm NN}}$ = 5.02\,TeV. Both plots are compared with MC event generators.}
\label{UE-pp-pPb}
\end{figure}
The underlying event is the collection of all particles not originating from the primary scattering or the related fragmentation. Conventionally, UE analyses include the charged-particle density measurement in the toward side (or near side, NS), transverse side (TS), and away side (AS) with respect to the leading trigger particle -- the highest transverse momentum ($p_{\rm T}^{\rm{leading}}$) track at midrapidity ($|\eta|<$0.8) of the collision. Left panel of Fig.~\ref{UE-pp-pPb} shows the charged-particle density as a function of $p_{\rm T}^{\rm{leading}}$ for TS in p--Pb collisions at $\sqrt{s_{\rm NN}}$ = 5.02\,TeV. A steep rise is seen in the event activity at low $p_{\rm T}^{\rm{leading}}$, and after $p_{\rm T}^{\rm{leading}}>$ 5 GeV/$c$, charged-particle density is nearly insensitive to the hard component as also observed in pp collisions~\cite{ALICE:2019mmy}. The figure also shows the comparison with PYTHIA8 (Angantyr)~\cite{Bierlich:2018xfw} and EPOS LHC~\cite{Pierog:2013ria} models. While PYTHIA8 describes the qualitative trend, the absolute level needs to be tuned. EPOS LHC, on the other hand, does not even describe the plateau behavior after $p_{\rm T}^{\rm{leading}}>$ 5 GeV/$c$. Right panel of Fig.~\ref{UE-pp-pPb} shows the charged-particle density as a function of $p_{\rm T}^{\rm{leading}}$ for pp and p--Pb collisions at $\sqrt{s_{\rm NN}}$ = 5.02\,TeV from AS with subtracted charged-particle density from TS to isolate the contributions from the fragmentation of the recoil jet. No strong difference between pp and p--Pb collisions at $p_{\rm T}^{\rm{leading}}>$ 8 GeV/$c$ is observed, which indicates no modification of the jet-like component in p--Pb collisions at the LHC. The results are also compared with PYTHIA8 (Monash)~\cite{Sjostrand:2014zea}, PYTHIA8 (Angantyr) and EPOS LHC models. As evident in the lower panel, the models fail to describe the ratio of charged-particle density from pp to p--Pb collisions at low $p_{\rm T}^{\rm{leading}}$. EPOS LHC describes qualitatively the downward dip at low $p_{\rm T}^{\rm{leading}}$, which suggests that the it could be related to flow.
\section{Very forward energy measurement in pp collisions at $\sqrt{s}$ = 13\,TeV}
\begin{figure}[ht!]
\centering
\includegraphics[scale=0.35]{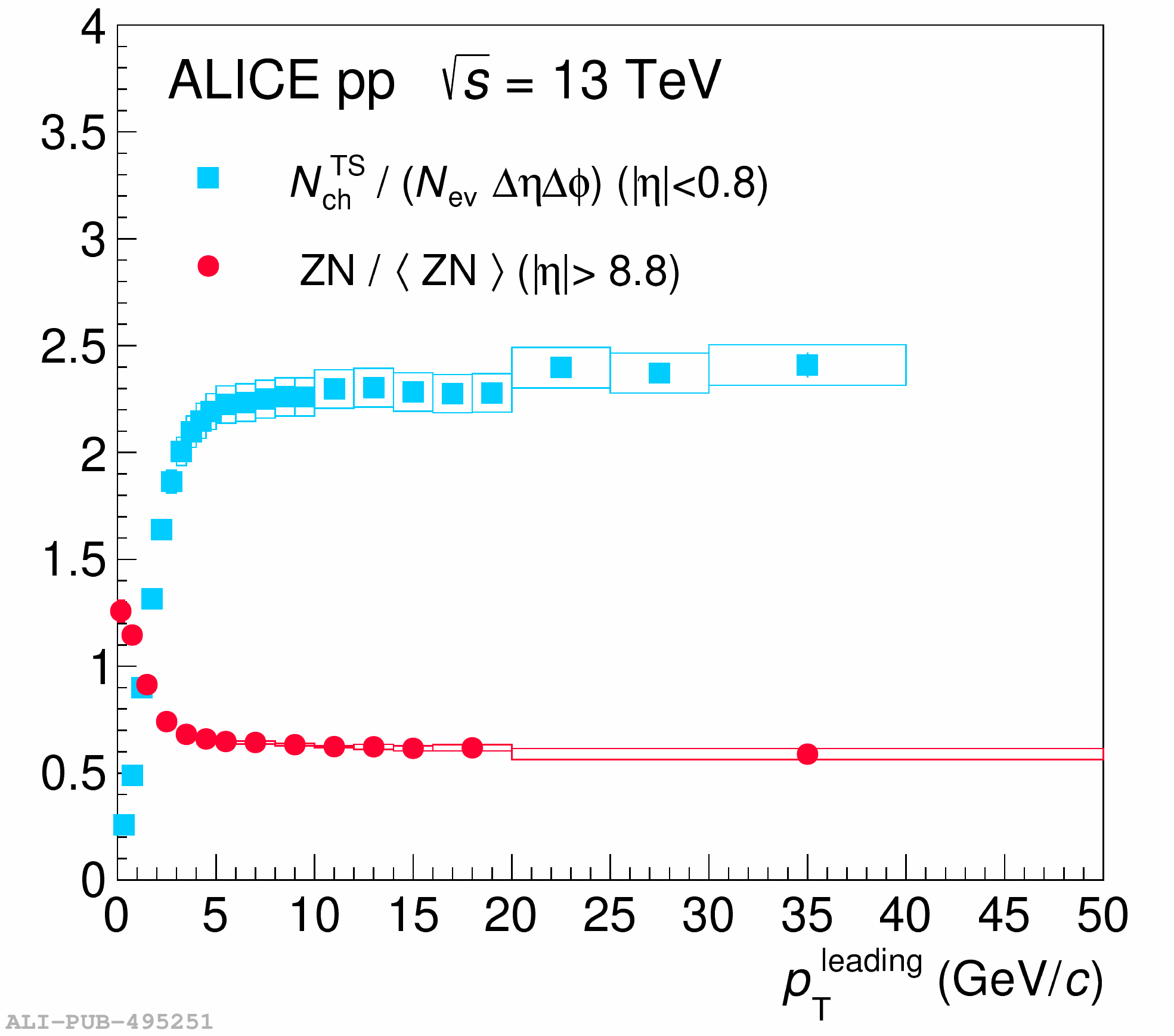}
\caption{Self-normalised ZN signal at forward rapidity and charged-particle density in the transverse region at midrapidity as a function of $p_{\rm T}^{\rm{leading}}$ in pp collisions at $\sqrt{s}$ = 13\,TeV}
\label{FwdEvsUE}
\end{figure}
The very forward energy measurement by the ZDC is a powerful tool to characterise the proton breakup, providing unique insights into the initial stages of the collision. Figure~\ref{FwdEvsUE} shows the self-normalised ZN signal (measured with the neutron calorimeter) at forward rapidity~\cite{ALICE:2021poe} and the charged-particle density in the transverse region at midrapidity~\cite{ALICE:2019mmy} as a function of $p_{\rm T}^{\rm{leading}}$ in pp collisions at $\sqrt{s}$ = 13\,TeV. The very forward energy decreases with increasing $p_{\rm T}^{\rm{leading}}$ and after 5 GeV/$c$ saturates. Similarly, the charged-particle density in TS rises as a function of $p_{\rm T}^{\rm{leading}}$ and then saturates at about the same $p_{\rm T}^{\rm{leading}}$. This indicates that the forward energy is correlated with midrapidity UE activity and that small forward energy detection selects events with high charged-particle multiplicity and high $p_{\rm T}^{\rm{leading}}$ particle at midrapidity.

\section{Strangeness production vs. effective energy in pp collisions at $\sqrt{s}$ = 13\,TeV}
\begin{figure}[ht!]
\centering
\includegraphics[scale=0.36]{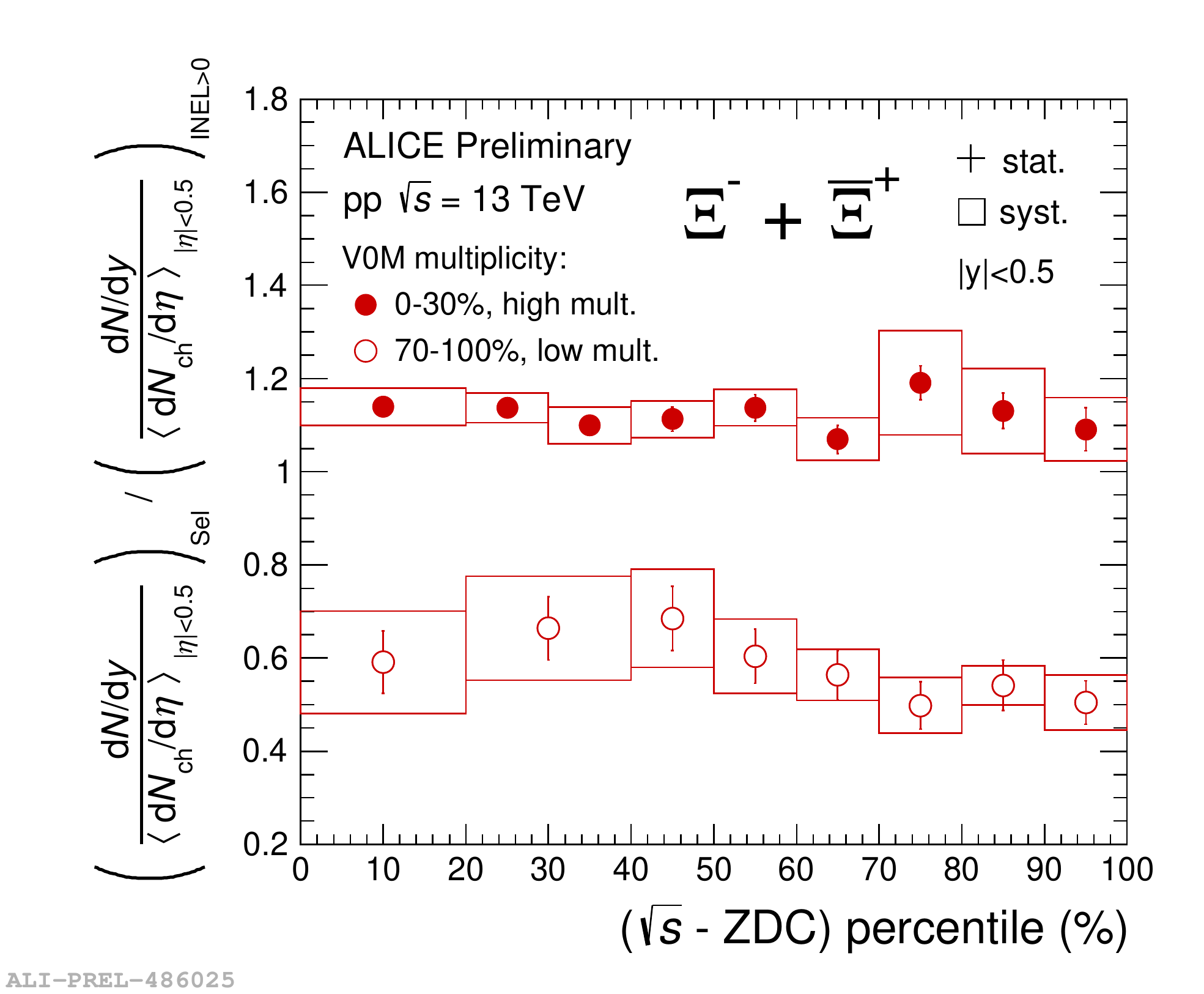}
\includegraphics[scale=0.35]{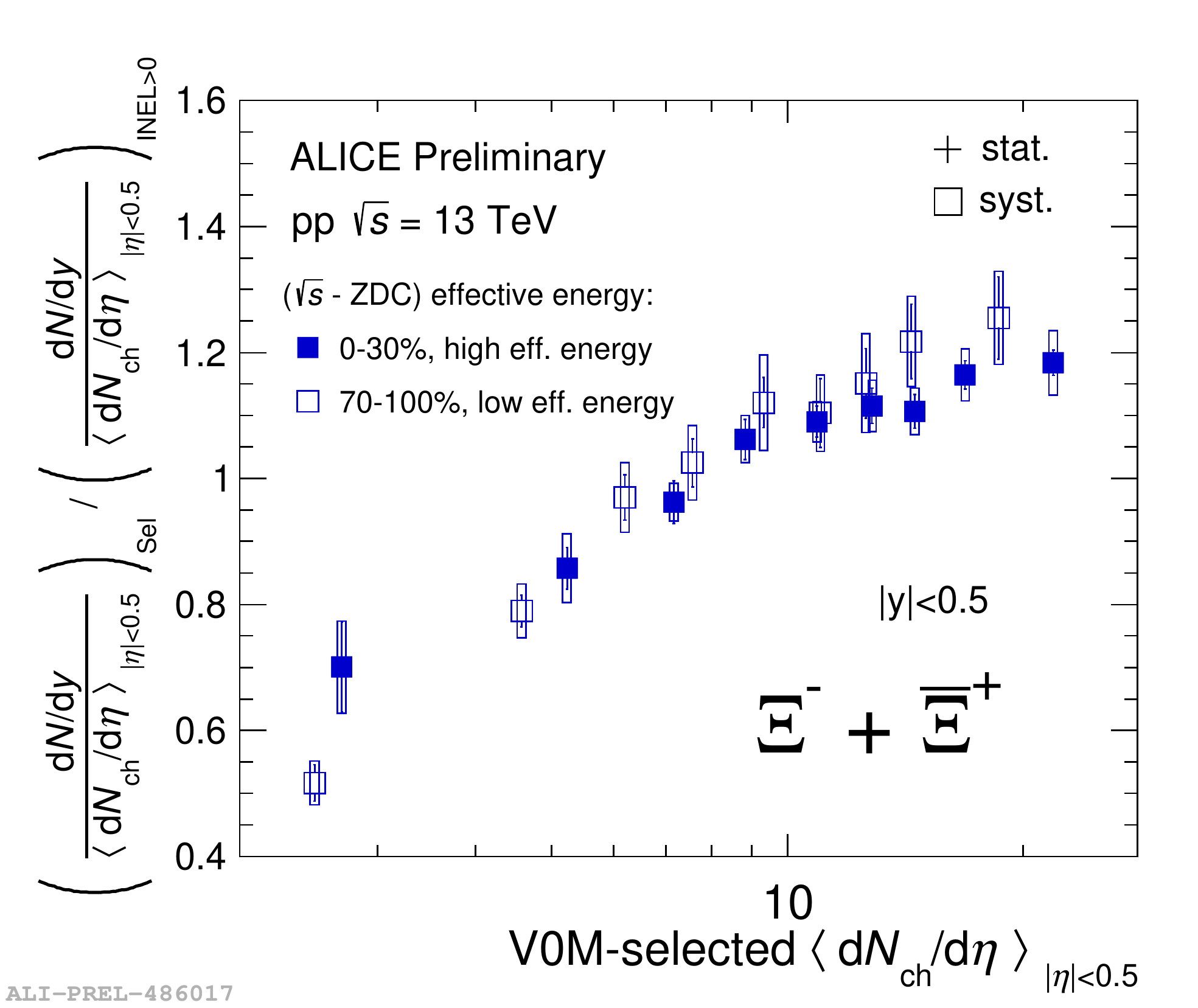}
\caption{Left (Right): self normalised ratio of integrated yield of $\Xi$ to charged-particle multiplicity selected using effective energy event classes (V0M multiplicity classes), fixing the multiplicity through the V0 estimator (effective energy event classes)}
\label{XiEffec}
\end{figure}
The part of the energy in the initial stage available for particle production is related to what we call here the effective energy $\sqrt{s} -$ E$_{\rm{ZDC}}$. In order to disentangle initial and final-state effects on strange hadron production, the events are classified in effective energy and charged-particle multiplicity classes using ZDC and V0 detectors, respectively. Left (Right) panel of Fig.~\ref{XiEffec} shows the self normalised ratio\footnote{The self normalisation is done with integrated yield of $\Xi$ to charged-particle multiplicity from INEL$>$0 events. Here, INEL$>$0 refers to the events with at least one charged particle in $|\eta|<1$. } of integrated yield of $\Xi$ to charged-particle multiplicity selected using effective energy event classes (V0M multiplicity classes\footnote{multiplicity classes through the V0 estimator}), fixing the V0M multiplicity classes (effective energy event classes). The ratio is found to be independent as a function of effective energy classes for fixed (0-30)\% and (70-100)\% V0M multiplicity classes. Similarly for fixed (0-30)\% and (70-100)\% effective energy classes, the ratio increases as a function of V0M multiplicity classes. The results suggest that the effective energy does not play a significant role in the strangeness enhancement observed in pp collisions but that the enhancement is mostly driven by the final-state charged-particle multiplicity.

\section{Summary}
We present the first UE measurements in p--Pb collisions at the LHC and find that PYTHIA8 (Angantyr) and EPOS LHC models fail to describe the experimental data. These results  provide useful input to tune the MC event generators. We observe no medium modifications of the away-side jet in p--Pb collisions at  $\sqrt{s_{\rm NN}}$ = 5.02\,TeV compared to pp collisions. Furthermore, we report the correlation between the forward energy measured by the ZDC at forward rapidity and the charge particle density at midrapidity. The very forward energy measurement provides information complementary to the measurements of the UE. Finally, using the forward energy information from the ZDC, a multi-differential study is shown to disentangle initial and final state effects on strange hadron production in pp collisions. It is found that the effective energy does not play a significant role in the strangeness enhancement observed in pp collisions and this suggests that this effect is mostly driven by the final state charged-particle multiplicity.

\end{document}